\documentclass[CEJP,PDF]{cej} 
\usepackage{layout}
\usepackage{amsmath}
\usepackage{textcomp}
\usepackage{hyperref}
\usepackage[pdftex]{graphicx}

\title{Asymmetric Double Quantum Wells with Smoothed Interfaces}

\articletype{Rapid Communication}

\author{Vladimir Gavryushin\email{v.gavriusinas@cablenet.lt} }

\institute{Institute of Applied Research and Semiconductor Physics department, Vilnius 
University,\\ Sauletekio al. 9-III, 10222 Vilnius, Lithuania }

\abstract{We have derived and analyzed the wavefunctions and energy states for an 
asymmetric double quantum wells, broadened due to static interface disorder 
effects, within well known discreet variable representation approach for 
solving the one-dimensional Schrodinger equation. The main advantage of this 
approach is that it yields the energy eigenvalues and the eigenvectors in 
semiconductor nanostructures of different shapes as well as the strengths of 
the optical transitions between them. We have found that interface 
broadening effects change and shift energy levels to higher energies, but 
the resonant conditions near an energy coupling regions do not strongly 
distorted. A quantum-mechanical calculations based on the convolution method 
(smoothing procedure) of the influence of disorder on the motion of free 
particles in nanostructures is presented.}


\begin{document}

\maketitle


\section{Introduction}

When a thin ($\sim $100 A) layers of one semiconductor (e.g. GaAs) are 
sandwiched between layers of another semiconductor with a larger bandgap 
(e.g. AlGaAs), carriers are trapped and confined in two dimensions (2D), due 
to the potential barriers. As a result of quantum confinement discrete 
energy states (or "subbands") occur, which change dramatically electronic 
and optical properties of such structures, called as the quantum wells 
(QWs). When the quantum wells are coupled there exist probabilities for the 
electron tunnel and be in either of the two wells. The novel optical 
properties of the 2D electron gas, associated with the transitions between 
quantized subbands, so called "intersubband transitions", correspond from 
mid-infrared to terahertz (THz) photon energies. They have narrow 
line-widths and extremely large transition dipole moments. 

In recent years, there has been considerable interest in asymmetrical 
multiple-quantum well systems, because many new optical devices based on 
intersubband transitions are being developed. This feature could fulfill the 
need for efficient sources of coherent infrared (IR) radiation for several 
applications, such as communications, radar, and optoelectronics. Their most 
spectacular applications are quantum well IR photodetectors [1] and the 
quantum cascade (QC) lasers [2] that relies on the intersubband transitions 
and resonant tunneling between adjacent QWs. Moreover, intersubband 
transitions give rise to extremely high linear and nonlinear optical 
susceptibilities. Specially engineered asymmetric quantum wells possess 
resonant 2$^{nd}$ and 3$^{rd}$ order optical nonlinearities which are 
respectively 3 and 5 orders of magnitude higher than that in the bulk 
semiconductors, which may be used for frequency mixing, phase conjugation 
and all-optical modulation [3]. These devices are made with epitaxially 
grown GaAs/AlGaAs and InGaAs/AlInAs [4]. With the recent 
development on semiconductor devices growth by the molecular beam epitaxy 
and metal-organic chemical vapor deposition techniques, multi-barrier 
quantum well structures are becoming the basic building blocks of modern 
semiconductor devices, such as resonant tunneling diodes (RTD) [5], 
far-infrared and THz lasers [6], quantum cascade (QC) lasers [2,7], etc.  

Real QW structures tend to deviate from the ideal homogeneous 
heterostructure with perfectly smooth interfaces. The reasons are the 
stochastic processes of the crystal growth leading to local variations of 
chemical composition, well width, and lattice imperfections to name a few. 
Since a QW is generally a heterostructure formed by a binary semiconductor 
(AB) and a ternary disordered alloy (AB$_{1-x}$C$_{x})$, as in 
InGaAs/GaAs, there are two types of disorders responsible 
for the inhomogeneous broadening: compositional disorder caused by 
concentration fluctuations in a ternary component and random diffusion 
across the interface. In order to design new devices or optimize the device 
performance, and thus properly predict their behavior, one needs to know the 
detailed information of quasi-bound levels in real disordered multi-barrier 
quantum well structures. Theoretical studies of effects due to compositional 
and interface disorders have a long history [8-10]. 

To understand the physical properties of the heterostructure devices, one 
needs to solve the eigenvalue problem of carriers in QWs. 
It is well known that exact analytic solutions to such problems 
are only available for simple structures such as square or parabolic well 
[11] and even in these structures, in general, in the presence of 
perturbations such as external fields, disorder effects [12], 
etc. the problem cannot be solved exactly. 

There have been various numerical methods used to 
calculate the band profiles in QWs: the matrix approach (MA) [13], the 
transfer matrix (TM) method [14,~15], the finite difference method (FDM) 
[16,~17], the finite element (FE) technique [18,~19], discreet variable 
representation (DVR) approach [20], envelope function (EF) method [21], 
Wentzel--Kramers--Brillouin (WKB) approximation [22], variational method 
(VM) [23], and Monte Carlo (MC) simulations [24]. Among them, the WKB and EF 
methods adopt approximations, thus give the results unreliable; the VM only 
works well at simple QWs and weak fields; the MC and FE methods are 
highly computer-orientated approaches; the MA usually require wave function 
to be well behaved. While DVR as also TM methods overcomes all the 
shortcomings listed above and could be easily applied to any potential 
profiles of biased/unbiased multi-barrier quantum well structures. 

In this paper we describe shortly a numerical technique based on the DVR 
approach, as a grid-point representation of a Hamiltonian matrix elements 
[12,~20,~25], which is capable of solving the eigenvalue and eigenfunction 
problems in an arbitrary QWs under arbitrary perturbation, for an asymmetric 
double quantum wells with interfaces broadened due to static disorder 
effects. 

\section{Calculation details}
\subsection{A model of the interface disorder effects}

Randomly distributed charged dopants, or composition x in mixed crystals 
A$_{x}$B$_{1-x}$, lead to unavoidable fluctuations of the doping impurities 
concentration on a microscopic scale. Two things influence as an interface 
disorder effects: coordinate fluctuations of interface position 
(Fig 1~a), and gap energy E$_{g}$(x) fluctuations 
(Fig 1~b,c) due to the randomly distributed 
dopants. These fluctuations result in potential fluctuations. This situation 
is schematically shown in Fig 1. The magnitude of 
band-edge energy fluctuations (Fig 1~b,c) caused by 
the random distribution of charged donors and acceptors was first calculated 
by Kane [26]. States with energy below the unperturbed conduction band edge 
or above the unperturbed valence band edge are called tail states, which 
significantly change the density of states in the vicinity of the band edge. 
The absence of order means that the wavevector \textbf{\textit{k}} is no 
longer a good quantum number. At energies high in the band this spread in k 
values can be described by a mean free path, but deep in the tail 
localization is complete. 

\begin{figure}
\includegraphics[scale=0.54]{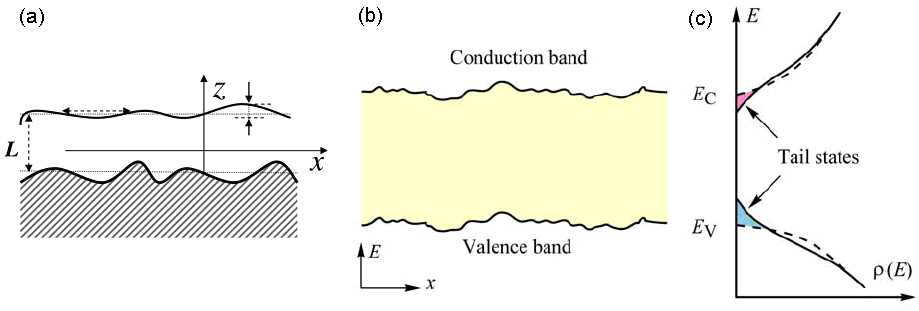}
\caption{(a) Characteristic length scales describing the interface roughness 
of a QW. (b) Spatially locally fluctuating band edges caused by random 
distribution of impurities (internal Franz-Keldysh effect), after [27]. (c) 
Resulting densities of states in the conduction with tail states extending 
into the forbidden gap. The dashed lines show the parabolic densities of 
states in undoped semiconductors.\label{fig1}}
\end{figure}

Particularly simple models of an electron moving in a random potential are 
possible. If the fluctuations are not too large, good approximation is 
obtained by calculating spectra for slightly different configurations and 
adding them up using some broadening weight factor. Inhomogeneous 
broadening, due to site variation produced by a random distribution of local 
crystal fields, results in a Gaussian type broadening [26]. Homogeneous 
broadening, from dynamic perturbations on energy levels and equally on all 
ions, leads to a Lorentzian type broadening. So, QW's barrier interface 
roughness may be approximated [12] by the \textit{convolution} of Heaviside step function $\Phi 
$(x-x$_{0})$ with an area normalized, moving Gaussian broadening envelope 
function of width $\sigma _{G}$: 
\begin{equation}
\label{eq1}
H_G (z,z_0 ,\sigma )=\frac{1}{\sqrt {2\pi } \sigma _G }\int\limits_0^\infty 
{\exp \left[ {-\frac{(z-x)^2}{2\sigma _G^2 }} \right]} \,\Phi (x-z_0 )dx,
\end{equation}
or with the Lorentzian broadening envelope function of width $\Gamma $: 
\begin{equation}
\label{eq2}
H_L (z,z_0 ,\Gamma )=\frac{\Gamma }{\pi }\int\limits_0^\infty 
{\frac{1}{(z-x)^2+\Gamma ^2}\,} \Phi (x-z_0 )dx.
\end{equation}
A convolution (smoothing) procedure [28] is an integral that expresses 
the amount of overlap of envelope function (i.e. Gaussian or Lorentzian) as 
it is shifted over another function $\Phi $. Instead of the $\infty $ limit 
in integrations usually is used any enough big value for the result 
convergation. The barrier steps may be broadened also in extremely simple 
analytical way, - by applying of the phenomenological atan(x) function 
against the Heaviside step function $\Phi $(x), usual for an ideal 
heterostructure with perfect interfaces: 
\begin{equation}
\label{eq3}
\Phi (z-z_i )\Rightarrow \frac{\pi }{2}+\arctan \frac{z-z_i }{\Gamma _i }
\end{equation}
where $\Gamma _{i}$ is broadening parameter of the interfaces at z$_{i}$. 
This function with zero mean z$_{i}$, characterize the deviation of the 
i$^{th}$ interface from its average position. Examples of calculated 
functions (\ref{eq1})-(\ref{eq3}) are presented in Fig 2. As we see 
in Fig 2, this function (\ref{eq3}) (red dots) may be a 
good approximation for convolution of Heaviside function by Lorentzian 
envelope (\ref{eq2}) (curve~3). 

\begin{figure}
\includegraphics[scale=0.52]{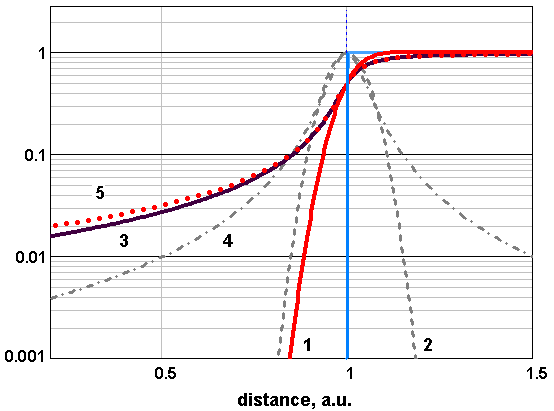}
\caption{Comparison of convolution broadenings of the Heaviside step-function 
with different broadening functions: 1 -- Gaussian broadened step (\ref{eq1}); 2 -- 
Gaussian envelope; 3 -- Lorentzian broadened step (\ref{eq2}); 4 - Lorentzian envelope; 
5 -- Analytical arctan(x) approximation curve (\ref{eq3}). Broadening parameter $\Gamma$~=~0.05~nm.
\label{fig2}}
\end{figure}

\subsection{Method of Discrete Variable Representation (DVR)}
 
Analytical expressions for asymmetrical double and triple quantum-wells 
are possible only for idealized rectangle QWs [29]. We select 
discrete variable representation (DVR) as a numerical method [20,~30] for 
our MathCAD calculations of stationary 1D Schrodinger equation for confined 
eigenstates. Different types of DVR methods have found wide applications in 
different fields of problems [31,~32]. We also carried out DVR calculations 
for QWs an unharmonic Morse potential [12,~25]. 

The DVR is a numerical grid-point method in which the matrix elements of the 
local potential energy operator V(r) is approximated as a diagonal matrix 
(mnemonic: DVR - \textbf{d}iagonal \textbf{V}(\textbf{r}), or 
\textbf{D}iscrete \textbf{V}ariable \textbf{R}epresentation): V$_{ik}$ = 
$\langle \phi _{i}\vert $V$\vert \phi _{k}\rangle =\delta 
_{ik}$V(x$_{i})$ [31], and the kinetic energy matrix is full, but it has 
simple analytic form, as a sum of 1D matrices. DVR method is selected 
since it avoid having to evaluate integrals in order to obtain the 
Hamiltonian matrix and since an energy truncation procedure allows the DVR 
grid points to be adapted naturally to the shape of any given potential 
energy surface. The DVR method greatly simplifies the evaluation of 
Hamiltonian matrix elements H$_{ik}$~=~$\langle \phi _{i}\vert 
$H$\vert \phi _{k}\rangle $ and obtains the eigenstates and 
eigenvalues by using standard numerical diagonalization methods of MathCAD 
or Mathematica. 

If we choose an equally spaced grid, x$_{i}$ = i$\Delta $x, (i = 
0,~$\pm $1,~$\pm $2,~... $\pm $N), then the DVR gives an extremely simple 
grid-point representation of the kinetic energy matrix $\hat {T}_{i,k} 
={\hbar ^2k_{i,k}^2 } \mathord{\left/ {\vphantom {{\hbar ^2k_{i,k}^2 } 
{2m^\ast }}} \right. \kern-\nulldelimiterspace} {2m^\ast }$ within the 
conditional formulation [20]: 
\begin{equation}
\label{eq4}
\mathord{\buildrel{\lower3pt\hbox{$\scriptscriptstyle\frown$}}\over {T}} 
_{i,k} =g\,\left\{ {{\begin{array}{*{20}c}
 {{\,\,\pi ^2} \mathord{\left/ {\vphantom {{\,\,\pi ^2} 3}} \right. 
\kern-\nulldelimiterspace} 3,{\begin{array}{*{20}c}
 \hfill & \hfill \\
\end{array} }} \hfill & {i=k} \hfill \\
 {\frac{2\,(-1)^{i-k}}{(i-k)^2},} \hfill & {i\ne k} \hfill \\
\end{array} }} \right\}.
\end{equation}
The only parameter involved being the grid spacing $\Delta $x via an 
energetically weighted grid parameter g (``energy quantum of the grid''): 
\begin{equation}
\label{eq5}
g=\frac{\hbar ^2}{2m^\ast }\left( {\frac{1}{\Delta x}} \right)^2
\end{equation}
where m$^{\ast }$ is the electron effective mass. So, if the grid points are 
uniformly spaced then numerical solutions of a matrix elements of the full 
energy Hamiltonian operator 
\[
\hat {H}=\mathord{\buildrel{\lower3pt\hbox{$\scriptscriptstyle\frown$}}\over 
{T}} +\hat {V}=-\frac{\hbar ^2}{2m^\ast }\frac{d^2}{dx^2}+V(x)\,
\]
is as [20]: 
\begin{equation}
\label{eq6}
\hat {H}_{i,k} 
=\mathord{\buildrel{\lower3pt\hbox{$\scriptscriptstyle\frown$}}\over {T}} 
_{i,k} +\hat {V}_{i,k} =\frac{\hbar ^2}{2m(\Delta x)^2}(-1)^{i-k}\left( 
{\frac{\pi ^2}{3}\delta _{i,k} +\frac{2\,}{(i-k)^2}(1-\delta _{i,k} )} 
\right)\,+V(x_i )\,\delta _{i,k} 
\end{equation}
when the $\delta $-functions are placed on a grid that extends over the 
interval x = (-$\infty $,$\infty )$. First term in parenthesis is a value 
of second term in the limit N~$\to $~$\infty $ [20]. 

In our calculations the potentials of an ideal and of a broadened double QW 
are used as [12]: 
\[
V_1 (x_i)=U_1 \,[\,1-\Phi (x_i +R_1 )+\Phi (x_i -R_1 )],
\]
\begin{equation}
\label{eq7}
V_2 (x_i )=V_1 (x_i )+U_2 \,[\,\Phi [x_i -(R_b +R_2 )]-\Phi (x_i -R)],
\end{equation}
and 
\[
V_1 (x_i )=U_1 \,[\,1+\frac{1}{\pi }\arctan {\kern 1pt}\left( {\frac{x_i 
}{\Gamma }} \right)-\frac{1}{\pi }\arctan {\kern 1pt}\left( {\frac{x_i +R_1 
}{\Gamma }} \right)],
\]
\begin{equation}
\label{eq8}
V_2 (x_i )=V_1 (x_i )+U_2 \,[\,\frac{1}{\pi }\arctan {\kern 1pt}\left( 
{\frac{x_i -(R_b +R_2 )}{\Gamma }} \right)-\frac{1}{\pi }\arctan {\kern 
1pt}\left( {\frac{x_i -R_b }{\Gamma }} \right)],
\end{equation}
correspondingly. Here $\Phi $(x) is a Heaviside step function; U$_{1(2)}$ 
are the depths of potential wells (differences in the offset band energies) 
for the 1$^{st}$ and 2$^{nd}$ QW of the widths R$_{1(2)}$; and R$_{b}$ is 
the barrier width. One can reasonably argue the restriction of used 
approximation for QWs disorder effects via an effective potentials (\ref{eq8}). 
Arguments may be directed to the problem being subject to short-range energy 
fluctuations which may be treated as a ``white-noise'' disorder with small 
correlation lengths. 

\section{Results and discussion}

Inter-well optical-phonon-assisted transitions are studied in an asymmetric 
double-quantum-well heterostructures [33] comprising one narrow and one wide 
coupled quantum wells (QWs). It is shown that the depopulation rate of the 
lower subband states in the narrow QW can be significantly enhanced thus 
facilitating the intersubband inverse population, if the depopulated subband 
is aligned with the second subband of the wider QW, while the energy 
separation from the first subband is tuned to the energy of optical phonon 
mode. 

Seeking to reproduce mentioned effects of [33], the eigenvalues (stationary 
energy states E$_{n})$ and the eigenvectors (wavefunctions $\psi _{n})$ 
for a quantum number $n$, as the solutions of the Schrodinger equation $\hat 
{H}\psi _n (r)=E_n \psi _n (r)$, were calculated using standard numerical 
diagonalization methods (eigenvals(H) and eigenvec(H) commands in MathCAD) 
for the DVR Hamiltonian (\ref{eq6}). 

Results of our DVR calculations of such a laser structure as in [33] are 
depicted in Fig 3 as the eigenstates together with 
the corresponding wavefunctions for the coupled asymmetric square quantum 
wells with the potentials of the form (\ref{eq7}). The states over the dissociation 
energy U are unbound and delocalized. Dependencies of the positions of the 
five lowest subbands B$_{i}$ as a function of the narrow well width values 
R$_{1}$ = 3, 5, 7, 10, 12, 15 nm for the fixed values of barrier width R$_{b}$ = 
2 nm and of the second QW width R$_{2}$ = 15 nm are shown in 
Fig 3. Broken lines are the same dependencies for 
the A$_{i}$ states of a 1$^{st}$, but single, quantum well with the same width 
R$_{1}$. Comparing the fans of dependencies for A$_{i}$ and B$_{i}$ states 
for single and double QW's, we can resolve the doublet nature of the states 
and especial coupling (anti-crossing) regions of the levels in double QW. 
The insets around in Fig 3 show the model band 
diagrams with energy levels and corresponding wavefunctions of an ADQW 
heterostructures (AlAs/GaAs). It is useful also to have some indication of how many grid points are 
necessary for this DVR to provide an accurate description of a quantum 
system. Convergence of the calculation can be checked by decreasing the 
number of grid used in calculation. For full convergence of the calculation 
results we found that enough number of grid is N $\ge $ 50, however we have 
used the number of calculus points N~=~500 for the better shaping of 
calculated wavefunctions. 

\begin{figure}
\includegraphics[scale=0.535]{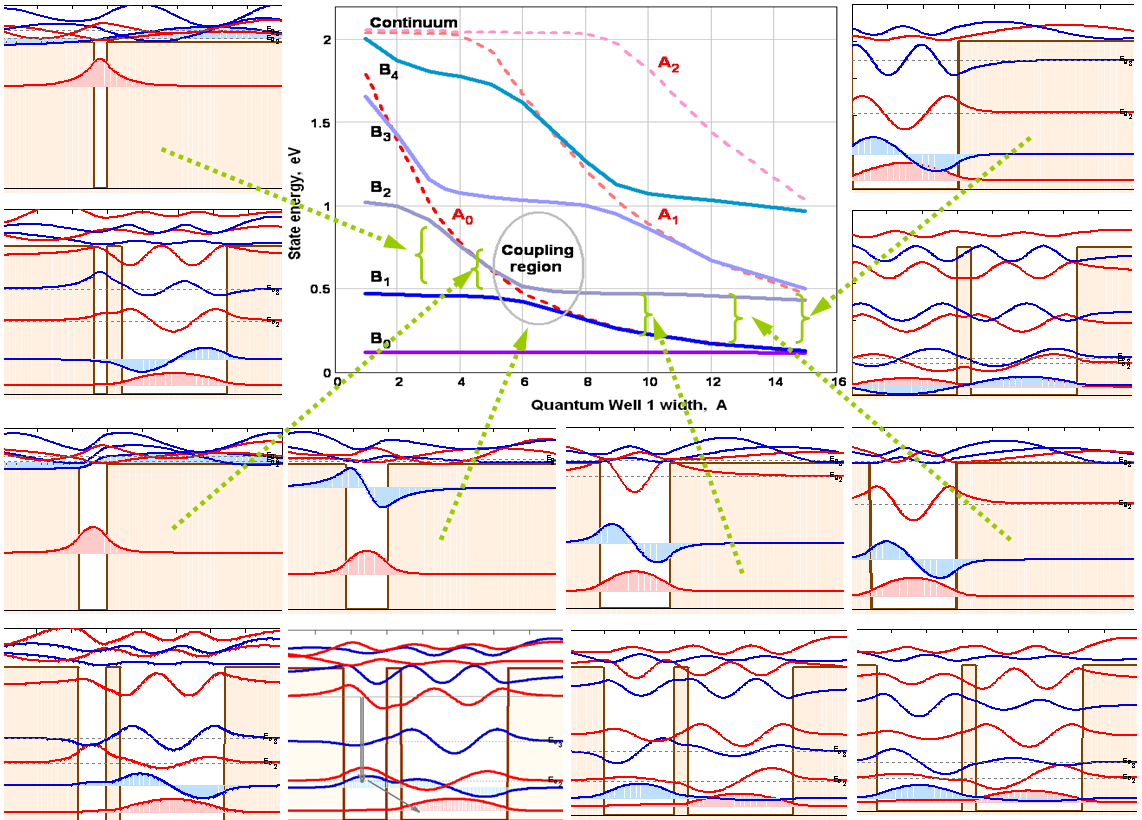}
\caption{Unbroadened ideal ADQWs case. Dependencies of the positions of the 
five lowest subbands B$_{0}$ -- B$_{4}$ (in eV) as a function of the 
narrower well width R$_{1}$ (= 3, 5, 7, 10, 12, 15 nm) for fixed values 
R$_{b}$ = 2 nm and R$_{2}$ = 15 nm. Broken lines are the same dependencies 
of the A states for a single quantum well of R$_{1}$ width. The insets 
around show the model band diagrams with energy levels and corresponding 
wavefunctions of an asymmetric double quantum well heterostructures. 
Double-lined arrow corresponds to the stimulating light-emitting transition, 
assisted by the resonant phonon emission (arrow between states n=1 and n=0) 
in the heterostructure. Calculated region: -30 nm $\div $ +30 nm with 500 
grids. \label{fig3}}
\end{figure}

All states in double QW are splitted doublets, 
because the degeneracy is unmounted by different parity properties. 
When the barrier thickness becomes smaller, quantum coupling due to the 
tunneling between the wells has the place. As a consequence, an energy 
splitting occurs (rounded regions indicated in Fig 3 and Fig 4) and the respective electron states, 
the so-called binding and anti-binding states, are delocalized over both 
wells. The energy splitting or tunnel coupling $\Delta $E is determined by 
the barrier thickness R$_{b}$ and height. The resonant situation can be 
obtained for asymmetric, coupled double quantum wells with applied bias 
[33]. The lowest coupled state is mainly localized in the wide well and the 
other state is mainly localized in the narrow well. Due to the coupling of 
the two wells the two states have nonzero probability density in both wells. 
Under a suitable bias these two states become resonant. 

\begin{figure}
\includegraphics[scale=0.4]{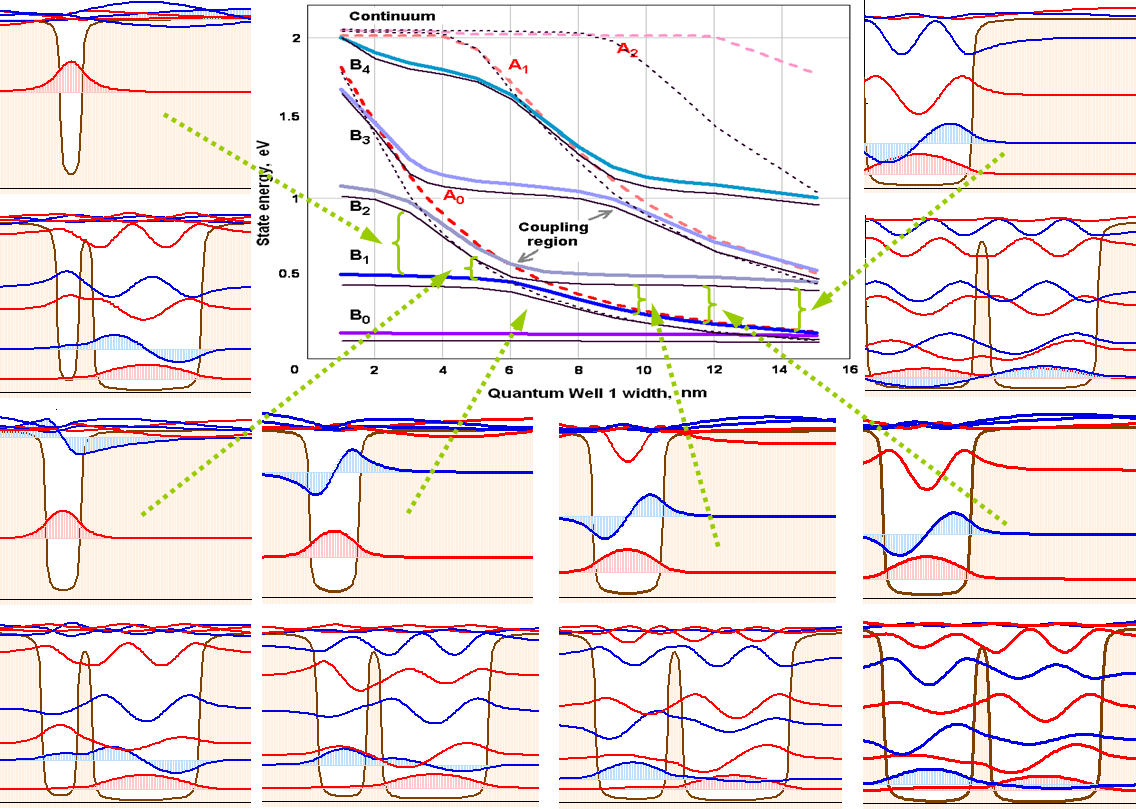}
\caption{Broadened ADQWs case. Dependencies of the positions of the five 
lowest subbands B$_{0}$ -- B$_{4}$ (in eV) as a function of the narrower 
well width R$_{1}$ (= 3,~5,~7,~10,~12,~15 nm) for fixed values R$_{b}$ = 
2~nm and R$_{2}$ = 15~nm. Broken lines are the same dependencies of the 
three A$_{i}$ states for a single QW with the width R$_{1}$. For comparison, 
thin black lines presents the case of ideally rectangle shaped ADQWs as in 
Fig 3. The insets around shows the model band 
diagrams with energy levels and corresponding wavefunctions. Interface 
roughness parameter $\Gamma $ = 0.2~nm. Calculated region: -30 nm $\div $ 
+30~nm with 500 grids. \label{fig4}}
\end{figure}

One embodiment of THz laser structure is depicted in Fig 3 and Fig 4 where the 
energy levels in coupling region are separated by energy equal to the 
optical phonon energy. Double-lined arrow corresponds to the stimulated 
light-emitting transition, assisted by the resonant phonon emission (arrow 
between states n=1 and n=0). Phonon-assisted transitions between the coupled 
levels depend strongly on the phonon energy involved in the transition. 

The results of DVR calculations of the same family of ADQWs as in Fig 3, but with structurally broadened interfaces 
with the potentials approximated by (\ref{eq8}), with used interface roughness 
parameter $\Gamma $ = 0.2~nm, are depicted in Fig 4. In the central part of Fig 4 both energy fans 
for an ideal rectangular interfaces (as in Fig 3) and 
for a broadened ones are presented together and mutually compared. Here the 
thin black lines presents the case of ideally rectangle shaped ADQWs. As we 
see, interface broadening change and shift energy levels to the higher energies, 
but the resonant conditions near an energy coupling regions do not strongly 
distorted. The blue shift is as a sequence of an effective narrowing of the distorted QWs, but the changes seen in the coupling regions are caused from the different inter-well barrier profile. 

\section{Conclusions}

The discreet variable representation approach for solving the 
one-dimensional Schrödinger equation is performed to calculate the 
structural and electronic properties of asymmetric double quantum wells 
broadened due to static interface disorder effects. We have derived and 
analyzed the wave functions, and the energy states for the structurally 
disordered (broadened) quantum wells of different profiles. The interlayer 
interactions would destroy the state degeneracy, induce more states, and 
vary their energy. Interface broadening effects change and shift energy 
levels to higher energies, but the resonant conditions near an energy 
coupling regions do not strongly distorted. 

\section*{Acknowledgments}

This work was partly supported by the Lithuanian State Science and Studies 
Foundation grand.

\end{document}